\documentclass[twocolumn]{aastex631}

\usepackage{orcidlink}

\PassOptionsToPackage{table}{xcolor}
\usepackage{xcolor}
\usepackage{bm}
\let\tablenum\relax
\usepackage{siunitx}
\usepackage{lineno}
\usepackage{xspace}
\usepackage[normalem]{ulem}
\DeclareSIUnit\parsec{pc}
\usepackage[table]{xcolor}
\usepackage{float}
\usepackage{multirow}
\usepackage{graphicx}
\usepackage{enumitem}

\shortauthors{\textnormal{Abdelmaguid et al.}}

\usepackage{hyperref}
\usepackage{upgreek} 
\usepackage{txfonts} 
\usepackage{float}

\usepackage{longtable, booktabs}

\submitjournal{ApJ}

\accepted{July 25, 2025}

\begin{document}

\title{The Timing Evolution of the Magnetar Swift~J1818.0$–$1607 During a Period of Reduced Activity}

\author[0000-0002-4441-7081]{Moaz Abdelmaguid} 
\affil{Department of Physics, New York University, 726 Broadway, New York, NY 10003, USA}
\affil{New York University Abu Dhabi, PO Box 129188, Abu Dhabi, UAE}
\affil{Center for Astrophysics \& Space Science (CASS), NYU Abu Dhabi, PO Box 129188, Abu Dhabi, UAE}

\author[0000-0003-1307-9435]{Paulo C. C. Freire}
\affil{Max-Planck-Institut für Radioastronomie, Auf dem Hügel 69, 53121 Bonn, Germany}

\author[0000-0003-4679-1058]{Joseph D. Gelfand}
\affil{New York University Abu Dhabi, PO Box 129188, Abu Dhabi, UAE}
\affil{Center for Astrophysics \& Space Science (CASS), NYU Abu Dhabi, PO Box 129188, Abu Dhabi, UAE}
\affil{Center for Cosmology and Particle Physics (CCPP, Affiliate), New York University, 726 Broadway, New York, NY 10003, USA}

\author[0000-0002-0862-6062]{Yogesh Maan}
\affil{ National Centre for Radio Astrophysics, Tata Institute of Fundamental Research, Pune 411007, Maharashtra, India}

\correspondingauthor{Moaz Abdelmaguid}
\email{m.abdelmaguid@nyu.edu}

\author[0000-0003-4136-7848]{Samayra Straal}
\affil{New York University Abu Dhabi, PO Box 129188, Abu Dhabi, UAE}

\author[0000-0002-2312-8539]{J. A. J. Alford}
\affil{New York University Abu Dhabi, PO Box 129188, Abu Dhabi, UAE}
\affil{Center for Astrophysics \& Space Science (CASS), NYU Abu Dhabi, PO Box 129188, Abu Dhabi, UAE}

\begin{abstract}

 We report results from an observational campaign of the radio-loud magnetar Swift~J1818.0$–$1607 using the Green Bank Telescope (GBT) at 2.0 GHz which began in 2021 November, during a period of reduced activity approximately 20 months after its 2020 March outburst. Over the $\sim 60$ day duration reported here, the integrated pulse profile remained consistently stable, exhibiting a single, narrow peak with a small precursor component and no evidence of a postcursor one.~This pulse profile is in sharp contrast to the double-peaked morphology observed during an observing campaign $\sim$ 120 days preceding ours. Along with this change in the integrated pulse profile shape, we also measure a slower spin-down rate ($\dot{\nu}$) compared to the end of that preceding campaign. Together, these differences suggest that a mode-switching event likely occurred between the end of that campaign and the start of ours. Finally, we derived a phase-connected timing solution from our data, from which we inferred a characteristic age of $\tau_{c} \sim2500$ years—about 2.5 times older than the most recent published estimate— and a surface dipole magnetic field strength of $B_{field} \sim1 \times 10^{14}$\,G, nearly three times weaker.These updated estimates reflect the short-term variations in the magnetar's spin-down rate, from which both its age and magnetic field strength are inferred, rather than intrinsic changes in the magnetar itself.

\end{abstract}


\section{Introduction} \label{sec:intro}

Magnetars are neutron stars believed to have extremely strong surface magnetic fields, ranging from $10^{13}$ to $10^{15}$ G, and typically have long spin periods (1 to 12 seconds) and small characteristic ages (e.g, \citealt{2017ARA&A..55..261K}, and the references therein). Unlike rotation-powered pulsars (RPPs), whose emission is driven by the loss of rotational energy, magnetars are believed to be powered by the decay of their ultra-strong magnetic fields. This idea was first proposed in the 1990s to explain the mysterious soft gamma repeaters \citep{1987ApJ...322L..21K} and anomalous X-ray pulsars,  whose X-ray luminosities ($L_X$) were found to far exceed their spin-down power ($\dot{E}$), implying an alternative energy source beyond rotational losses \citep{1992ApJ...392L...9D}.

\begin{table*}[htbp]
\centering
    \caption{Summary of past \& present monitoring campaigns of Swift~J1818.0$–$1607 including this work}
\begin{tabular}{cccc}
\toprule
\toprule
MJD & Telescope  & Frequency (GHz) &  Reference \\
    \midrule
58922 -- 59015 & $*$ & 1.37 -- 6  & \citep{2020MNRAS.498.6044C}\\
58936 -- 59092 & TMRT & 2.24 -- 8.69 & \citep{2021MNRAS.505.1311H}\\
58977 -- 59128 & Parkes & 2.4  & \citep{2020ApJ...896L..37L}\\
59053 -- 59105 & Lovell \& MK II & 1.53 & \citep{2022MNRAS.512.1687R}\\
59092\ -\ 59190 & GBT & 0.8\ --\ 35 & \citep{2025arXiv250215200L} \\
59104 -- 59365 & Lovell \& MK II & 1.53 & \citep{2024MNRAS.528.3833F}\\
59117 -- 59400 & Effelsberg & 6.0 & \citep{2022MNRAS.512.1687R}\\
59520 -- 59578 & GBT & 2.0 & This Work \\
\bottomrule
\end{tabular}%
\tablecomments{\footnotesize { TMRT: Shanghai Tian Ma Radio Telescope, GBT: Green Bank Telescope\\
$^*$ Observations reported by \cite{2020MNRAS.498.6044C} were made using many telescopes; Efflsberg at 1.37, 2.55, 4.85 \& 6 GHz, Lovell at 1.53 GHz and Nançay Radio Telescope (NRT) at 1.48 GHz.  MK II is the Mark II radio telescope located at Jodrell Bank Observatory.
}}
\label{tab:monitoring}%
\end{table*}%

Magnetars exhibit a wide range of observational transient phenomena, including short (0.1 s), bright X-ray bursts, and giant flares that last for seconds to minutes in duration longer. These outbursts are usually accompanied by a change in their spectral and timing behavior \citep{2017ARA&A..55..261K}. They also show significant variability in their spin-down rates, along with discrete glitching events and changes in pulse profile shapes that point to a highly dynamic and evolving magnetosphere \citep{2021MNRAS.502..127L, 2022MNRAS.512.1687R}.

Of the $\sim$30 currently identified magnetars, pulsed radio emission has only been detected in only six of them \citep{2014ApJS..212....6O}. The appearance of this radio emission appears to be correlated with outbursts observed at higher energies \citep{2006Natur.442..892C, 2007ApJ...666L..93C} and typically features significant flux and spectral variations \citep{2020ApJ...896L..37L}. Additionally, the pulsed radio emission of magnetars display integrated profiles that vary within hours to days, along with a relatively flat radio spectra, characterized by spectral indices $\alpha > -\, 0.8$ \citep{2015MNRAS.451L..50T, 2017MNRAS.465..242T}, where the flux density $S_{\nu} \propto \nu^{\alpha}$. This is in contrast to RPPs, which usually have stable integrated profiles and steeper radio spectra, with typical values of $\alpha \sim -\, 1.6$  \citep{2018MNRAS.473.4436J}.

Swift~J1818.0$-$1607 was discovered in 2020, following its bright X-ray outburst detected by the \emph{Swift}--Burst Alert Telescope (BAT) \citep{2004ApJ...611.1005G}. Subsequent observations with the Neutron Star Interior Composition Explorer (NICER) revealed X-ray pulsations with a period of 1.36 s \citep{2020ATel13551....1E, 2020ApJ...896L..30E}, establishing it as the magnetar with the shortest known period to date. Shortly afterwards, radio pulsations were detected by the Efflesberg telescope \citep{2020ATel13553....1K}, making it the fifth known ``radio-loud'' magnetar. The initial observed spin-down rate suggests  a characteristic age of $\tau_{c} \sim$ 250 $–$ 1000 years \citep{2020MNRAS.498.6044C}, implying that Swift~J1818.0$-$1607 is one of the youngest known magnetars.

Similar to other radio-loud magnetars, Swift~J1818.0$–$1607 shows significant variability in its pulse profile morphology, spin-down rate, and flux across a range of timescales \citep{2025arXiv250215200L, 2024MNRAS.528.3833F, 2022MNRAS.512.1687R, 2021MNRAS.505.1311H, 2020ApJ...896L..37L, 2020MNRAS.498.6044C, 2020ApJ...902....1H}.~Furthermore, it has displayed numerous distinct emission modes, reflecting a highly dynamic and evolving magnetosphere. During its 2020 outburst, \citep{2021MNRAS.502..127L} identified two such emission modes below 4 GHz that alternated on minute-long timescales. Subsequent monitoring at 1.4 GHz by \cite{2022MNRAS.512.1687R}, along with simultaneous observations at 2.2 and 8.5 GHz by \cite{2021MNRAS.505.1311H} and \cite{2023MNRAS.523.2401B}, revealed additional mode-switching episodes with pulse profiles that differed markedly from the initial two. These observations further emphasize the magnetar’s complex emission behavior across multiple frequencies. Table~\ref{tab:monitoring} summarizes previous radio monitoring campaigns of this source, including the observations presented in this work.

Given the magnetar's dynamic nature, continued long--term monitoring of Swift~J1818.0$–$1607 is essential for a comprehensive understanding of the later stages of magnetar evolution. In particular, observations during periods of reduced activity provide an opportunity to explore whether magnetar magnetospheres settle into a stable configuration post-outburst, or continue to evolve over longer timescales.

Here, we present radio observations of Swift~J1818.0$–$1607 with the 100-m Green Bank Telescope (GBT), during a period of reduced activity.~We concentrate on a subset of our broader monitoring campaign, conducted between 2021 November and 2021 December, during which we could derive a phase-connected timing solution.~In \S\ref{sec:radio_obs}, we present the timeline of our observations and describe the data reduction procedures. The results are presented in \S\ref{sec:analysis_and_results}, including the timing solution (\S\ref{sec:timing_solution}), pulse profile evolution (\S\ref{sec:pulse_profile_evolution}), flux density and spectral indices measurements (\S \ref{sec:flux_density_results}). In \S\ref{sec:discussion}, we compare these findings with earlier studies on this source and other magnetars, and discuss their implications.~We finally conclude with a summary and future outlook in \S\ref{sec:summaryandconclusions}.

\section{Observations and Data Reduction} \label{sec:radio_obs}

\begin{figure}
    \centering
    \includegraphics[width=1.05\linewidth]{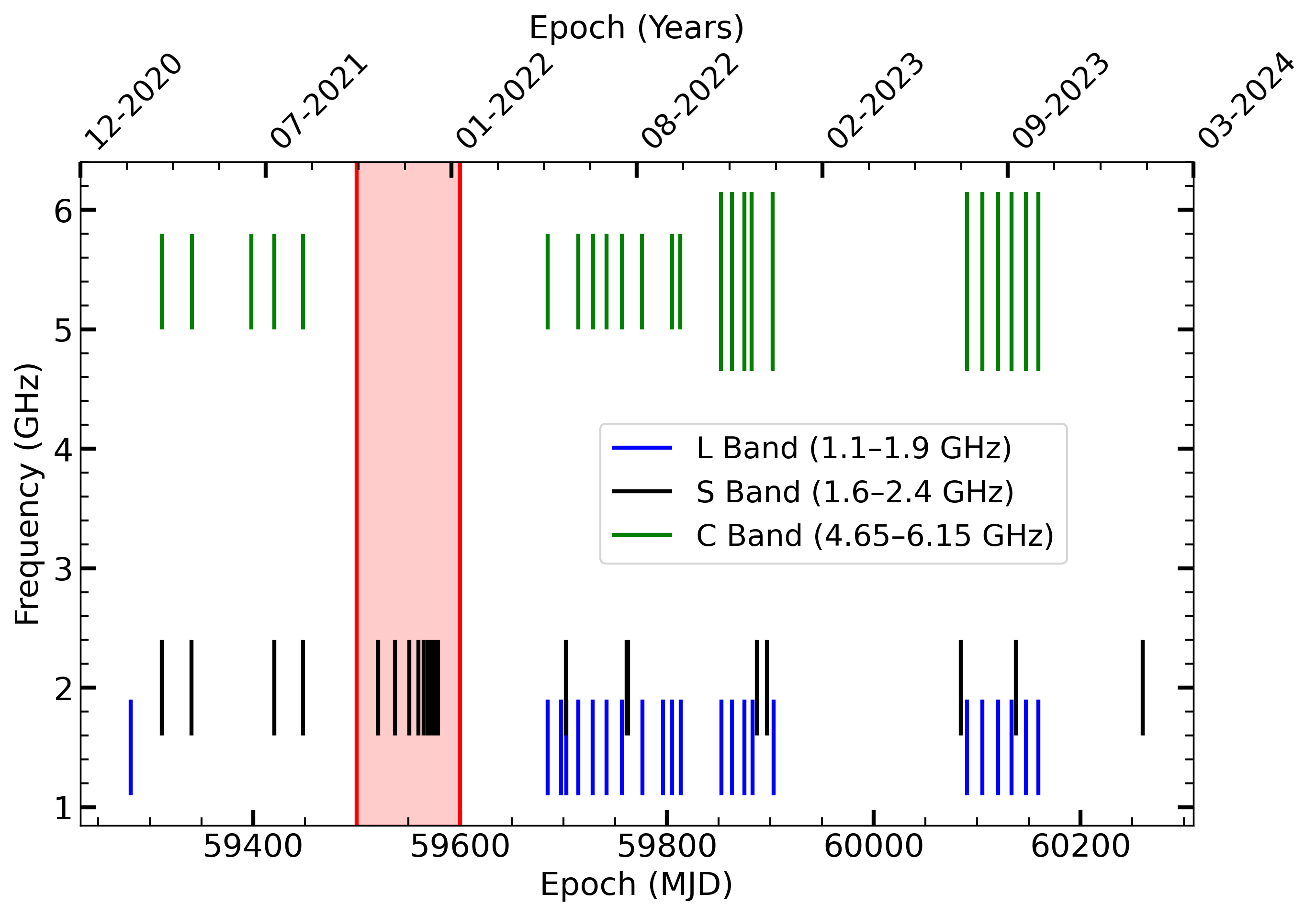}
    \caption{A summary of our entire monitoring campaign highlighting the dense--set presented in this paper.
}
    \label{fig:j1818_observations}
\end{figure}

We observed Swift J1818.0–1607 between 2021 March 18 (MJD 59291) and 2023 November 12 (MJD 60260) with the GBT in three frequency ranges; L band (1.1 $-$ 1.9 GHz), S band (1.6 $-$ 2.4 GHz) $\&$ C band (4.65 $-$ 6.15 GHz). 
Figure \ref{fig:j1818_observations} shows a timeline of our entire multi-frequency monitoring campaign.~The measurements from the full set of observations will be presented in future work; this paper focuses on the closely spaced S band-only observations conducted between 2021 Nov 2 (MJD 59520) $\&$ 2021 Dec 30 (MJD 59578), henceforth referred to as the ``dense set''  (2021B semester; Project code GBT21B-354, PI: Samayra Straal).

These observations were carried out in SEARCH mode, and the data were recorded using the VErsatile GBT Astronomical Spectrometer (VEGAS) backend \citep{2015ursi.confE...4P} with a bandwidth of 800 MHz at a central frequency of $\nu$ = 2 GHz and a \SI{40.96}{\micro\second} time resolution. The 800 MHz bandwidth was divided into 2048 frequency channels for better radio frequency interference (RFI) mitigation and more accurate dispersive delay correction. Each epoch included a \SI{65}{\second} obsevations of 3C286 used for absolute flux and polarization calibration.~A log of the dense set observations showing relevant details is shown in Table \ref{tab:observations}.

We analyzed the raw data from each epoch using the pulsar analysis software \texttt{PRESTO} \citep{2002AJ....124.1788R}. Below, we outline the detailed steps used to extract the times of arrival (ToAs) at each epoch, which are then used to construct the phase-connected timing solution described in \S\ref{sec:timing_solution}:

\begin{enumerate}[itemsep=0.0pt] 
    \item Automatically removing RFI using the \texttt{rfifind} routine in \texttt{PRESTO},
    \item Folding the data from each epoch with the \texttt{prepfold} routine, using the timing solution derived by \citet{2022MNRAS.512.1687R} as the initial ephemeris.
    \item Use the \texttt{pygaussfit.py} routine to generate a template pulse profile by fitting a Gaussian to the highest S/N observation, which will then be correlated with the folded profiles from each epoch.
    \item Use the \texttt{get\_toas.py} routine to extract at least 3 ToAs per epoch.
\end{enumerate}

\begin{table}[htbp]
\centering
    \caption{Log of the S band dense-set observations highlighted in red in Figure \ref{fig:j1818_observations} }
\begin{tabular}{cccc}
\toprule
\toprule
MJD & Date & Duration  & $F_{c}$ \\
- & (YYYY MM DD) & (Minutes)    & (GHz) \\ 
    \midrule
59520.76 & 2021 Nov 02 & 15 & 2.0  \\
59536.71 & 2021 Nov 18 & 15  & 2.0 \\
59550.88 & 2021 Dec 02  & 15  & 2.0 \\
59559.84 & 2021 Dec 11 &  15  & 2.0 \\
59564.80  & 2021 Dec 16 & 15  & 2.0 \\
59568.72   & 2021 Dec 20 & 20  & 2.0  \\ 
 59572.63  &2021 Dec 24& 15 & 2.0 \\
59576.59   & 2021 Dec 28 & 11.4  & 2.0\\
 59577.76  & 2021 Dec 29 & 24.92  & 2.0\\
 59578.80  &  2021 Dec 30& 15 & 2.0\\
\bottomrule
\end{tabular}%
\tablecomments{ $F_{c}$ denotes the central observing frequency. All observations were conducted with a bandwidth of 800 MHz.}
\label{tab:observations}%
\end{table}


We used on-source and off-source scans of 3C286 to flux calibrate the data, following the procedure outlined below:

\begin{enumerate}
    \item Using the \texttt{DSPSR} software library \citep{2011PASA...28....1V}, we fold the data using the phase-connected timing solution presented in Table \ref{tab:timing_solution}. This results in an "archive" file. 
    \item Using the \texttt{pazi} tool in the \texttt{PSRCHIVE} software package \footnote{https://psrchive.sourceforge.net/}\citep{2004PASA...21..302H}, we manually removed all RFI to minimize baseline variations through the following steps:

Scrunch the updated archive'' file in frequency to reduce the number of channels, then use the \texttt{pazi} routine to identify and flag RFI-affected sub-integrations. Use the resulting \texttt{psrsh} command to apply these flags back to the original, full-resolution archive file. This produces an archive'' file cleaned of RFI in both frequency and time domains.

    \begin{itemize}
        \item Scrunch the data in time and use the \texttt{pazi} routine to remove the frequency channels contaminated by RFI and then generate a \texttt{psrsh} command to print-out a script to reproduce this interactive session results and finally apply this to the original ''archive'' file from 1.
        \item Scrunch the updated ``archive'' file from the previous step in frequency and use the \texttt{pazi} routine to remove the RFI-affected sub-integrations. Use the resulting \texttt{psrsh} command to the original .ar file. This produces an ``archive'' file cleaned of RFI in both frequency and time domains. 
        \item Remove RFI from the ON \& OFF calibrator scans as well as the calibrator scan of the noise diode using the \texttt{pazi} routine. 
    \end{itemize}
    \item Using the \texttt{PSRCHIVE} software package \citep{2004PASA...21..302H}, we followed the calibration procedure outlined by \cite{2012AR&T....9..237V} to estimate the system equivalent flux density (SEFD). Here is a brief summary of the steps:
    \begin{itemize}
        \item Create a calibrator database that has the ON and OFF scans of the calibrator source (3C286) using the \texttt{pac} routine. 
        \item Make the ``.fluxcal'' file and update the created database by adding this file to it using the \texttt{fluxcal} routine. 
    \end{itemize}
    \item Check that the derived calibrator solutions ``.fluxcal'' file is free from RFI. If not, use \texttt{pazi} to remove contaminated channels.
    \item Calibrate the data using the derived calibrator solution --``.fluxcal'' file-- from the calibrator database.
\end{enumerate}

\section{Analysis and Results } \label{sec:analysis_and_results}

\subsection{Timing Solution} \label{sec:timing_solution}

\begin{table}[tbp]
\centering
    \caption{Timing parameters of Swift J1818.0$-$1607 derived from the observations listed in Table \ref{tab:observations}  }
\begin{tabular}{cc}
\toprule
\toprule
Parameter & Value \\
    \midrule
Right Ascension (J2000) (hh:mm:ss)& 18:18:00.23\\
Declination  (J2000) (dd:mm:ss)&   --16:07:53.00\\
DM (\si{\parsec\per\centi\metre\cubed}) &  710 $\pm$ 1\\
Date Range (MJD) & 59536.713 -- 59578.804\\
Epoch of Frequency &  59564.805772 \\
F0 (Hz)&  0.7326046915(5) \\
F1 (\si{\hertz\per\second}) & $-\, 4.4855(11)\, \times$ $10^{-12} $ \\
F2 (\si{\hertz\per\second\squared}) & $ 1.2286(85)\, \times$ $10^{-19} $ \\
EPHEM       &        DE200 \\
\bottomrule
\end{tabular}%
\tablecomments{ Both the DM and the the coordinates of the magnetar were held fixed. }
\label{tab:timing_solution}%
\end{table}%

\begin{figure}
    \centering
    \includegraphics[width=1.1\linewidth]{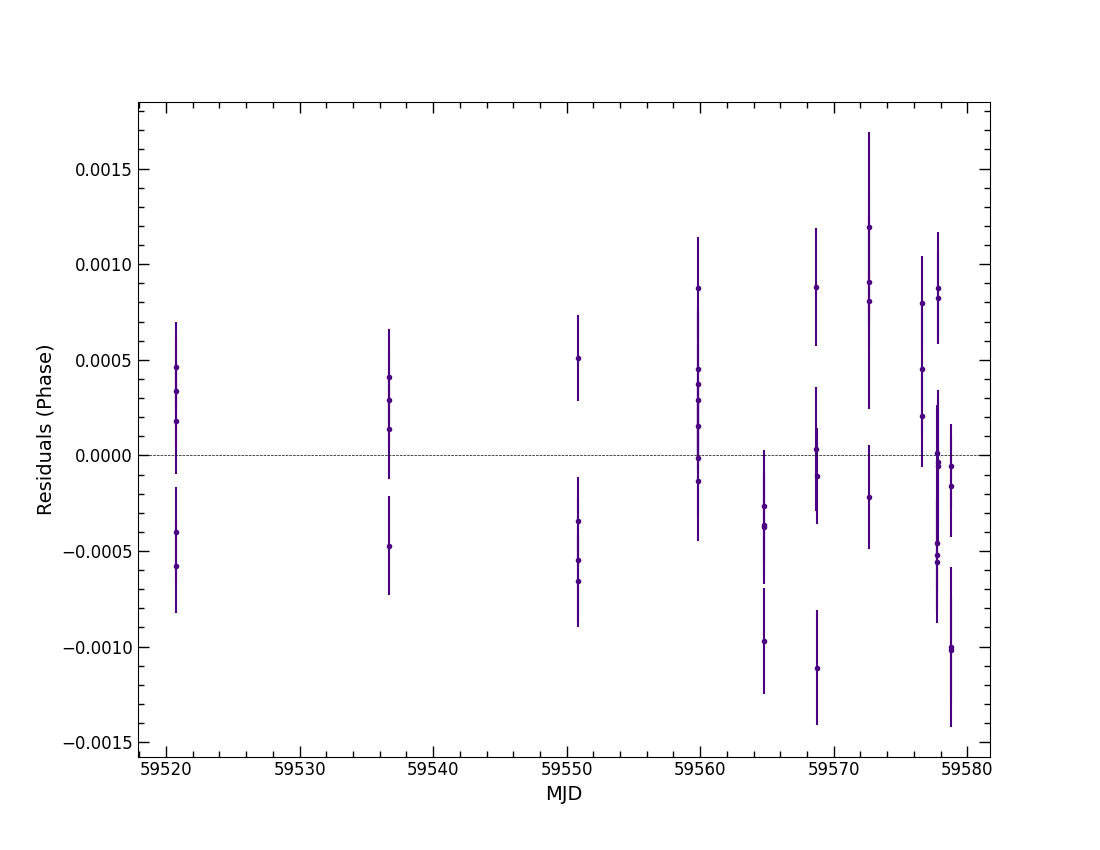}
    \caption{Timing residuals for Swift~J1818.0$-$1607 from the GBT observing campaign at 2.0 GHz. The reference MJD is 59564.8.}
    \label{fig:residuals}
\end{figure}

With the extracted ToAs, obtained using the procedure outlined in \S\ref{sec:radio_obs}, we used the \texttt{TEMPO2} pulsar timing software \citep{2006ChJAS...6b.189H} to derive a phase-connected timing solution following the technique described by \citet{2018MNRAS.476.4794F}. We kept the RA and DEC values fixed at the best-known X-ray coordinates reported by \citet{2020ApJ...904L..19B}, as was  done  by \citet{2022MNRAS.512.1687R}\footnote{This position was adopted because our timing analysis was conducted\\ in 2022, prior to the publication of the updated VLBI position from\\ \cite{2024ApJ...971L..13D}.}. Since the magnetar was observed at a single frequency band during the dense set, constraining the DM proved difficult.~Each observation was initially folded using \texttt{PRESTO}, allowing for a DM search around the value reported in the most recent timing solution by \citet{2022MNRAS.512.1687R}. The resulting folds consistently yielded a DM of $710 \pm 1\ \text{pc}\ \text{cm}^{-3}$, in agreement with the value reported by \citet{2022MNRAS.512.1687R}. We therefore adopted this value and fixed the DM in subsequent steps.~The derived phase coherent ephemeris is listed in Table \ref{tab:timing_solution} and the timing residuals are shown in Figure \ref{fig:residuals}. We were unable to extend the timing solution to later or earlier epochs due to significant timing noise, which made maintaining a phase-connected solution increasingly difficult.

\begin{figure*}
    \centering
    \includegraphics[width=1.0\linewidth]{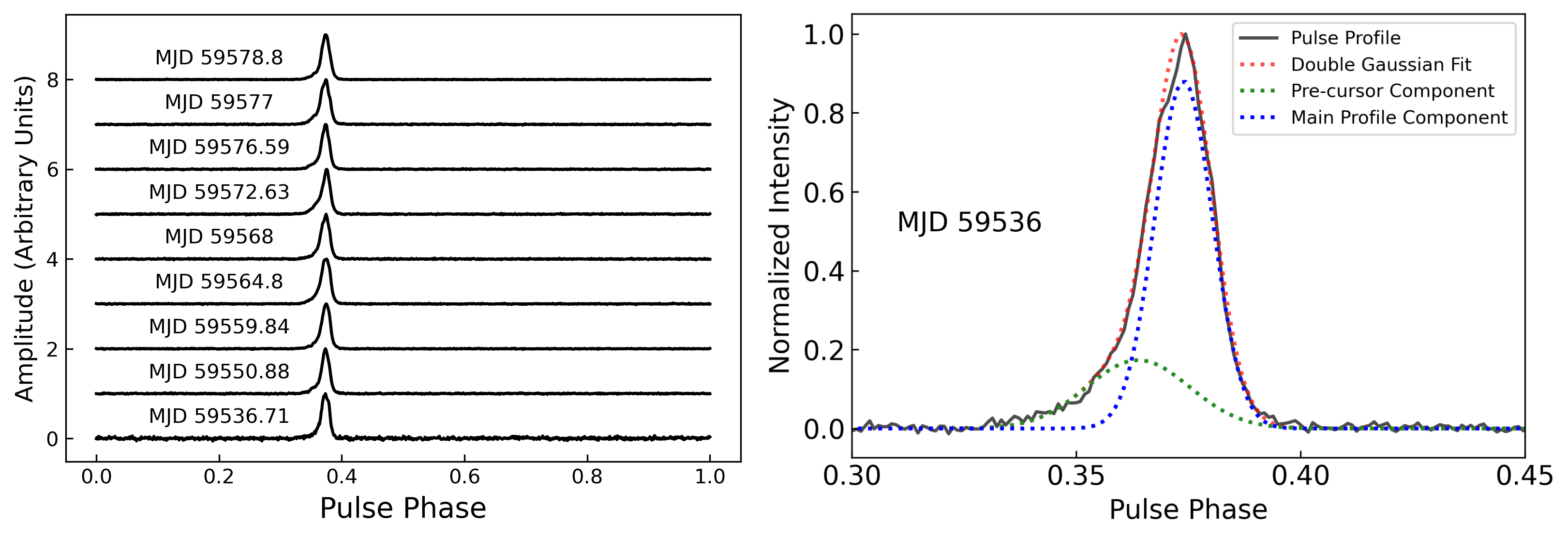} 
    \caption{Pulse Profile Evolution at 2.0 GHz over our monitoring period of $\sim$ 60 days. All profiles have been normalized for easier comparison. \textit{Right Panel:} An example of a double gaussian fit to the pulse profile at MJD 59536, showing the contributions from a narrow main component (blue dotted curve) and a broader pre-cursor component (green dotted curve). The red dotted curve represents the sum of these individual components.
}
    \label{fig:pulse_profile_evolution}
\end{figure*}

\begin{table}[htbp]
\centering
\caption{Pulse width, flux density, and spectral index measurements for Swift J1818.0$-$1607 during our monitoring campaign at 2.0 GHz.}
\begin{tabular}{cccc}
\toprule
\toprule
MJD & Pulse Width (FWHM) & $S_{\nu}$ &  $\alpha$ \\
- & (ms)    & (mJy)  &  - \\ 
    \midrule
59520.76 & 18.6(8) & 0.25(3) &  $-0.37(9)$  \\
59536.71 & 18.9(2) & 0.21(2)  &     $-0.15(16)$  \\
59550.88 &  18.6(8)   & 0.27(3)   & $-0.17(31)$ \\
59559.84 &  20.0(2)  &  0.27(3)      & $-0.20(20)$ \\
59564.80  & 22.6(8) & 0.23(2)   &  $-0.30(17)$ \\
59568.72   & 21.3(5) & 0.27(3)    &  $0.24(24)$ \\ 
59572.63  & 18.6(8) & 0.21(2)    & $-0.77(8)$\\
59576.59   & 18.6(8) & 0.33(3)   & $0.33(23)$ \\
 59577.76  & 22.6(8) & 0.26(3)     & $0.72(30)$ \\
 59578.80  & 20.0(2) & 0.31(3) &    $-0.10(23)$  \\
\bottomrule
\end{tabular}%
\tablecomments{ The table lists the fitted full-width at half-maximum (FWHM) pulse widths, period-averaged flux densities ($S_{\nu}$) at 2.0 GHz, and spectral indices ($\alpha$) at each observing epoch. Values are quoted with $1-\sigma$ uncertainties in parentheses, representing the uncertainty in the last digit(s).}
\label{tab:fluxdensity_pulse_width}%
\end{table}%


\begin{figure*}
    \centering
    \includegraphics[width=1.0\linewidth]{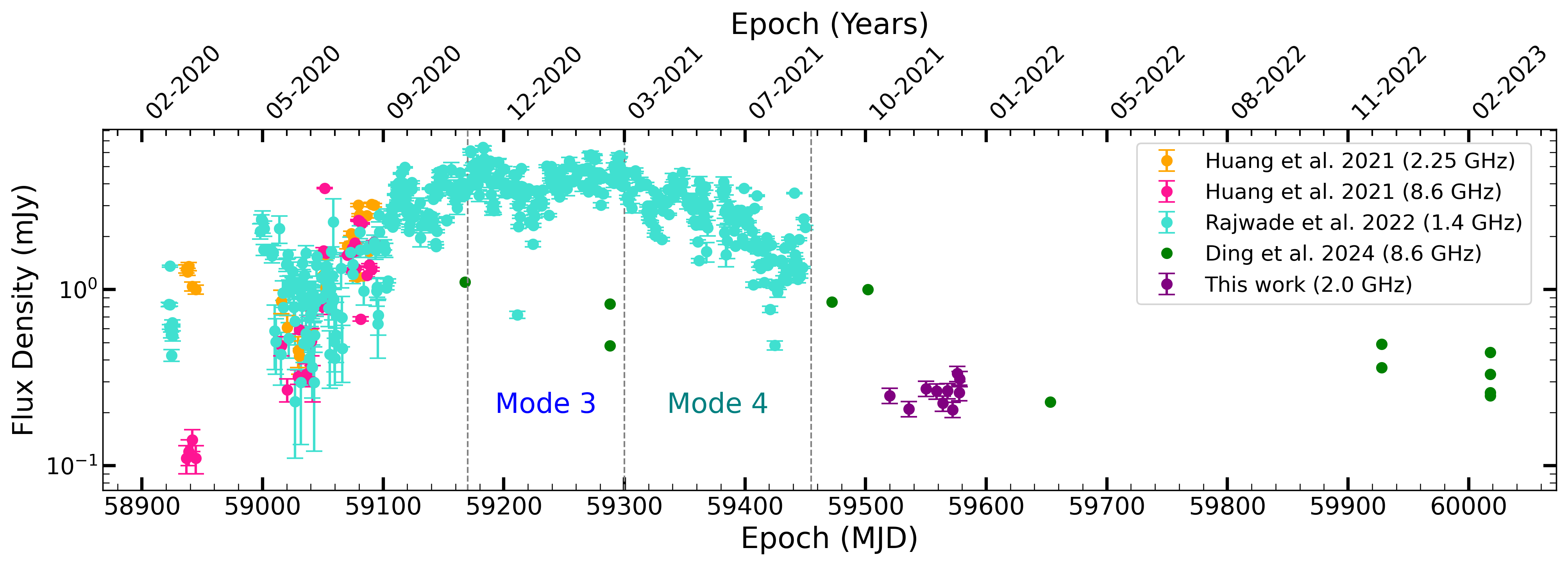}
    \caption{Period averaged flux densities as a function of MJD. The y-axis is shown on a logarithmic scale. Our 2.0 GHz measurements are plotted alongside earlier observations at 2.25 \& 8.6 GHz by \cite{2021MNRAS.505.1311H} and at 1.4 GHz by \cite{2022MNRAS.512.1687R} and future epochs at 8.6 GHz by \cite{2024ApJ...971L..13D}. The segments labeled Mode 3 and Mode 4 correspond to distinct mode-switching events, as identified by \citet{2022MNRAS.512.1687R}. }
    \label{fig:flux_density_evolution}
\end{figure*}

\begin{figure*}
    \centering
    \includegraphics[width=1.0\linewidth]{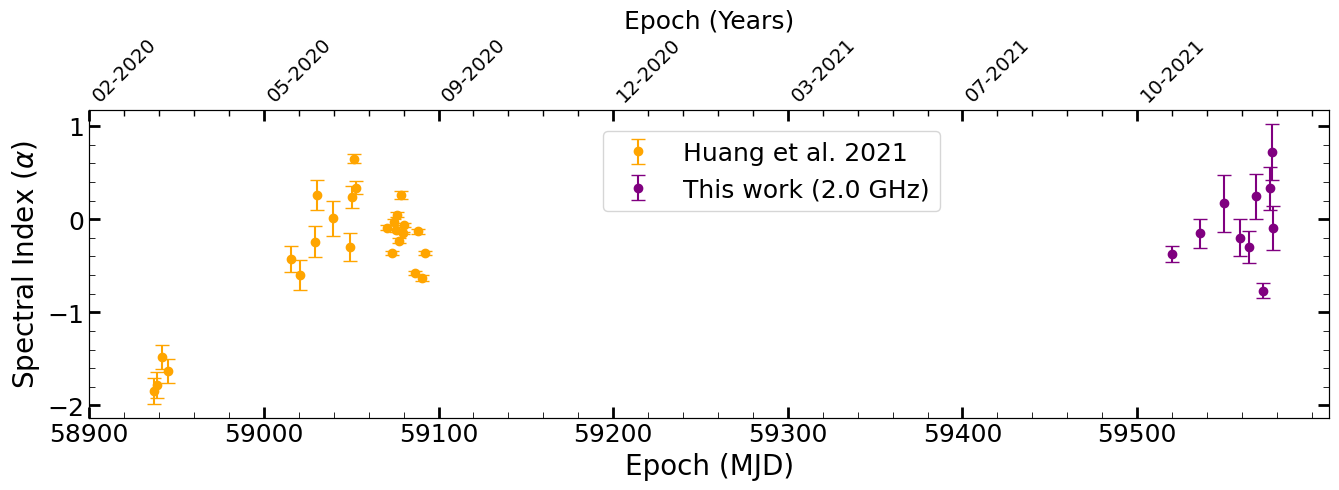}
    \caption{Power--law spectral indices as a function of MJD estimated using the period averaged flux densities along with older data reported by \citet{2021MNRAS.505.1311H}}
    \label{fig:spectral_index_evolution}
\end{figure*}

\subsection{Pulse Profile Evolution} \label{sec:pulse_profile_evolution}

Using the updated ephemeris derived in \S\ref{sec:timing_solution} and shown in Table \ref{tab:timing_solution}, we refolded the data at each epoch to obtain the corresponding integrated pulse profile. Left panel of Figure \ref{fig:pulse_profile_evolution} shows the temporal evolution of the pulse profile for the dense$-$set presented in this paper ($\sim$ 60 days). To make comparison easier, the profiles have been normalized.~The profiles are characterized by a dominant single-peak structure with an excess on the leading edge, indicative of a precursor component.

To better characterize the asymmetric morphology, and accurately measure the full-width half-maximum (FWHM or $W_{50}$), we fit each profile using a two-component Gaussian model.~The primary narrow Gaussian (blue dotted curve in Figure \ref{fig:pulse_profile_evolution}) accounts for the main pulse component, while the broader, low-amplitude secondary Gaussian (green dotted curve in Figure \ref{fig:pulse_profile_evolution}) captures the pre-cursor component.~The model was fitted using \texttt{scipy.optimize.curve\_fit}, which uses the  non-linear least-squares minimization method.~As shown in Figure~\ref{fig:pulse_profile_evolution}, this model captures the profile shape more effectively than a single Gaussian.~The resulting pulse widths, listed in Table~\ref{tab:fluxdensity_pulse_width}, range from $W_{50} \sim$ 18 $–$ 23 ms and exhibit minimal variation across epochs, indicating a stable pulse morphology over the monitoring period.~The fitted parameters for each Gaussian component are summarized in Table~\ref{tab:gauss_fit_parameters}.

\vspace*{\fill}

\subsection{Flux Density and Spectral Index} \label{sec:flux_density_results}
Following the calibration procedure outlined in \S \ref{sec:radio_obs}, we estimated the period-averaged flux density by dividing the data from each epoch into four sub-bands and integrating the area under the flux-calibrated profiles of each sub-band followed by dividing by the number of bins.~The overall period-averaged flux density across the entire frequency band was then determined by averaging the flux density values from the four sub-bands. The random errors on our flux density estimates are tiny, however, the flux density of our primary calibrator 3C286 itself could have uncertainties of the order of 5\% \citep{PB17}. Moreover, the presence of some faint residual RFI could also affect the measurements.~So, conservatively, we assume the corresponding flux density uncertainites to be 10\%.

The period-averaged flux density values each epoch are listed in Table \ref{tab:fluxdensity_pulse_width}. Our measurements indicate a relatively low flux density for the source, ranging between $S_{\nu} \sim 0.2-0.3$ mJy. Figure \ref{fig:flux_density_evolution} shows a plot of the evolution of our flux density measurements with time along with measurements from earlier epochs by \cite{2021MNRAS.505.1311H} \& \cite{2022MNRAS.512.1687R} and some future epochs by \cite{2024ApJ...971L..13D}. This will be discussed later in \S \ref{sec:mood_switching}.

\begin{figure*}
    \centering
    \includegraphics[width=1.0\linewidth]{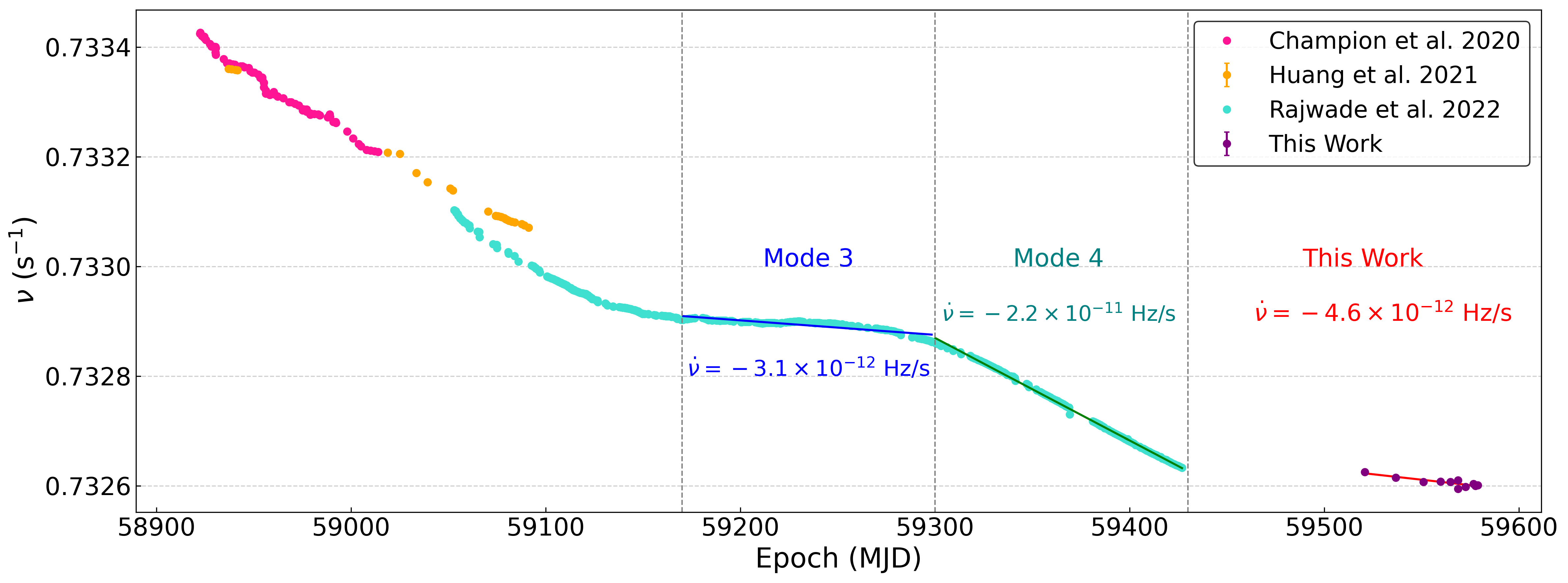}
    \caption{Spin frequency evolution of Swift~J1818.0$-$1607 over time, showing measurements from different studies \citep{2020MNRAS.498.6044C, 2021MNRAS.505.1311H, 2022MNRAS.512.1687R}, and this work. The segments labeled Mode 3 and Mode 4 correspond to distinct spin-down states associated with mode-switching events as identified by \cite{2022MNRAS.512.1687R}, each with a characteristic frequency derivative ($\dot{\nu}$) indicated by the fitted slopes. Data from \citet{2021MNRAS.505.1311H} was provided by Zhi-Peng Huang, who also shared the dataset from  \citet{2020MNRAS.498.6044C}, which had been extracted using the engauge digitizer app. Data from \cite{2022MNRAS.512.1687R} was provided by Kaustubh Rajwade. }
    \label{fig:spin_frequency_slopes}
\end{figure*}

Finally, given the large bandwidth of 800 MHz in the GBT's S-band receiver, we estimated the in-band spectral index at each epoch by fitting a power$-$law model to the measured flux densities across the 4 sub-bands above.~We performed this fit using the non-linear least squares optimization method available through \texttt{scipy.optimize.curve\_fit}. We assumed that, at a given frequency $\nu$, the flux density, $S_{\nu}$ follows the relationship $S_{\nu} \propto \nu^{\alpha}$. The corresponding spectral index values ($\alpha$) at each epoch are provided in Table \ref{tab:fluxdensity_pulse_width}, and its evolution over time is illustrated in Figure \ref{fig:spectral_index_evolution} along with $\alpha$ measurements at earlier epochs \citep{2021MNRAS.505.1311H}. Our spectral index measurements indicate that Swift~J1818.0$-$1607 continues to exhibit a flat radio spectrum, with  $\alpha \gtrsim -\, 1$, consistent with previous results reported by \citet{2021MNRAS.505.1311H}, particularly between MJD 59015 and 59090, as shown in Figure \ref{fig:spectral_index_evolution}.

\section{Discussion} \label{sec:discussion}

\subsection{Spin-down Behaviour \& Mode Switching} \label{sec:mood_switching}

Mode switching refers to the abrupt transition between two or more distinct emission states, thought to arise as a result of changes in the neutron star's magnetosphere \citep{2010MNRAS.408L..41T, 2007MNRAS.377.1383W}.~Swift~J1818.0$-$1607 has exhibited multiple distinct emission modes, reflected in both short- and long-term variations in its pulse profile. Between 2020~June and 2021~August, mode switching was observed at 1.4~GHz \citep{2022MNRAS.512.1687R} and simultaneously at 2.2 and 8.5~GHz \citep{2021MNRAS.505.1311H, 2023MNRAS.523.2401B}. On minute timescales, \citet{2020ApJ...896L..37L} reported two mode-switching behaviors below 4~GHz, characterized by morphologies distinct from those seen at higher frequencies. Since the closest published observations to our dataset are those of \citet{2022MNRAS.512.1687R}, we compare our results to this work. In their study, they identified distinct emission states, including the segments labeled Mode 3 and Mode 4 (Figures \ref{fig:flux_density_evolution} \& \ref{fig:spin_frequency_slopes}). These modes were characterized by differences in pulse profile morphology, spin-down behaviour and flux density. Crucially,  \cite{2022MNRAS.512.1687R} noted that these transitions between different modes or "mode switches" were correlated with a change in $\dot{\nu}$ (see Figure 2 in \citealt{2022MNRAS.512.1687R}), similar to what is observed for other mode-changing pulsars \citep{2010Sci...329..408L}. 

While different from the mode-switching seen in Swift J1818.0$-$1607, the radio-loud magnetar PSR~J1622$-$4950 showed another form of coupled spin and profile evolution, marked by a steady decline in its spin-down rate from 2011 to 2014, accompanied by a gradual narrowing of its pulse profile \citep{2017ApJ...841..126S}. However, other radio magnetars do not show clear evidence of a similar secular decrease in profile width. Additionally, the magnetar's spin-down rate reached a minimum just before the source became radio-silent, rising again when pulsations reactivated \citep{2017ApJ...841..126S, 2018ApJ...856..180C}.
In contrast, XTE~J1810$-$197 displayed significant pulse profile variability throughout its active phase but no evidence of a systematic or gradual narrowing \citep{2016ApJ...820..110C, 2019MNRAS.488.5251L}. Similar to PSR~J1622$-$4950, its spin-down rate also reached minimum values just before radio silence and increased again upon reactivation \citep{2016ApJ...820..110C, 2019MNRAS.488.5251L}.

A related behavior has been observed in PSR~J1119$-$6127, a high magnetic field pulsar with magnetar-like characteristics. Following its outbursts in 2016, an increase in the spin-down rate was observed to be correlated with an increase in its radio flux density \citep{2018MNRAS.480.3584D}. Altogether, the correlated changes in spin-down rate and radio emission observed in Swift~J1818.0$-$1607 are not anomalous, but rather align with trends seen in other magnetars and high B-field pulsars.

As mentioned earlier and shown in Figure \ref{fig:pulse_profile_evolution}, the integrated pulse profiles throughout our $\sim$ 60-day observing campaign consistently feature a single, narrow peak with little variation in shape, that closely resembles the pulse profile seen in emission mode 3 (\citealt{2022MNRAS.512.1687R}; Figure 3c).~In contrast, the closest published pulse profile to our data (near MJD 59400, \citealt{2022MNRAS.512.1687R}, Figure 3d) displayed a double-peaked structure with a precursor component. This indicates that the magnetar has since transitioned to a single-peak profile and remained in that state throughout our  monitoring period.

To compare the spin--down behavior of Swift~J1818.0$-$1607 across different epochs, we first measured the spin frequency at each individual epoch in our dataset. We then fitted linear slopes to the spin frequency data for Mode 3 (MJD 59170 $–$ 59300) and Mode 4 (MJD 59300 $–$ 59426), as well as for our own data spanning MJD 59520 $–$ 59578. The comparison shown in Figure \ref{fig:spin_frequency_slopes}  shows that the spin-down rate observed in our dataset ($\dot{\nu} = -\, 4.6 \times 10^{-12}\,  \mathrm{Hz\,s^{-1}}$) is notably similar to that of Mode 3 ($\dot{\nu} = -\, 3.1 \times 10^{-12}\,  \mathrm{Hz\,s^{-1}}$), suggesting a likely return to a similar emission state following the more rapid spin-down episode seen in Mode 4 ($\dot{\nu} = -\, 2.2 \times 10^{-11}\,  \mathrm{Hz\,s^{-1}}$).

Concurrently, our flux density measurements show that Swift~J1818.0$-$1607 remained in a persistently low-flux state with minimal variability throughout our observations (Figure \ref{fig:flux_density_evolution}). This overall behaviour closely resembles the relatively steady trend seen in Mode 3, albeit at systematically lower flux levels, and contrasts with the declining flux characteristic of Mode 4 observed immediately prior to our campaign. It is important to note that the flux density measurements shown in Figure \ref{fig:flux_density_evolution} span a range of observing frequencies (1.4 $-$ 8.6 GHz). Since magnetar emission can exhibit frequency-dependent behavior, these spectral variations may introduce biases when comparing flux measurements taken at different epochs.~This pattern of highly variable spin-down rates is not exclusively limited to Swift~J1818.0$-$1607.~For example, 1E~1048.1$-$5937 exhibited repeated increases and decreases in its spin-down rate—by up to an order of magnitude—within $\sim$ 100 $–$ 600 days following its multiple X-ray outbursts in 2002, 2007, 2012 \citep{2015ApJ...800...33A}. Following its 2016 outburst, 1E~1048.1$-$5937 showed a monotic decline in the magnitude of its spin-down variations; however, its long-term spin-down evolution remains dominated by recurrent fluctuations \citep{2020ApJ...889..160A}.

1E~1547.0$-$5408, another magnetar, underwent a rapid spin-down rate increase after its 2008, 2009 and 2022 outbursts, with significant variability persisting thereafter \citep{2012ApJ...748....3D, 2012ApJ...748..133K, 2023ApJ...945..153L}. Similar variability in the spin-down behaviour has also been observed in PSR~J1622$-$4950 \citep{2017ApJ...841..126S, 2018ApJ...856..180C}, and in XTE~J1810$-$197 \citep{2016ApJ...820..110C, 2019MNRAS.488.5251L, 2023ApJ...956...93H}.~Additionally, the high-magnetic-field pulsar PSR~J1846$-$0258 showed erratic spin evolution during its 2020 outburst, including a spin-up followed by a steady decline in spin-down rate \citep{2024ApJ...976...56S}.~These examples demonstrate that spin-down variability is a common feature among magnetars in the aftermath of outbursts, especially when they are observed over long timescales.~Thus, the variability observed in Swift~J1818.0$-$1607 appears consistent with the broader behavior of magnetar population. However, given that only a comparatively short time has passed since its outburst in 2020, continued long-term monitoring will be essential to determine whether it follows similar evolutionary trends.

Therefore, taken together; the change in pulse profile, spin$-$down rate ($\dot{\nu}$) and flux density between Mode 4 and the onset of our observations suggests that another mode switching episode likely occurred during the $\sim$100-day gap (MJD 59426 to MJD 59520) between the last published timing campaign \citep{2022MNRAS.512.1687R} and the start of ours. This underscores the importance of continued monitoring of Swift~J1818.0$–$1607 to capture potential future transitions or changes in its activity, given the magnetar's highly dynamic and evolving behavior.

 \subsection{Pulse Widths}

In our study of Swift~J1818.0$-$1607,  we observed pulse widths ranging from $W_{50} \sim 18-23$ ms) at 2.0 GHz, implying a stable integrated profile shape throughout the observation period. This is in contrast with the variability reported immediately following the magnetar's outburst, where the pulse profile transitioned from a broader ($W_{50} \sim 40-80$ ms) to narrower one ($W_{50} \sim 10-45$ ms) at (1.37 $-$ 2.55) GHz \citep{2020MNRAS.498.6044C}. Additionally, \cite{2025arXiv250215200L} reports a mostly consistent and comparatively narrower pulse profile widths ($W_{50} \sim 16-18$ ms)  at higher frequencies between 6 and 22 GHz (MJD 59092 $-$ 59190). This narrowing of pulse width at higher frequencies is consistent with what has been observed in many RPPs \citep{2016A&A...586A..92P} and was also observed in XTE~J1810$-$197 \citep{2021PASJ...73.1563E}.  In canonical pulsars, the evolution of pulse width with frequency is explained by radius-frequency mapping, which predicts that the pulse profile becomes narrower at higher radio frequencies \citep{1978ApJ...222.1006C}.

\begin{table*}[htb]
\centering
    \caption{A comparison of the characteristic age ($\tau_{c}$), spin-down inferred magnetic field ($B_{field}$) and spin$-$down luminosity ($\dot{E}$) estimated in this paper with values reported in previous studies and observational campaigns.}
\begin{tabular}{ccccc}
 \toprule
\toprule
 Reference & MJD &  $\tau_{c}$  & Spin$-$down Inferred $B_{field}$ &  $\dot{E}$ \\
    \midrule
\cite{2020ATel13559....1C} & 58922 &  $\sim 265$ yr & $ \sim 3.4 \times 10^{14} \, \text{G}$  &  \(\sim  1.6 \times 10^{36} \ \mathrm{erg \, s^{-1}} \) \\
\cite{2020MNRAS.498.6044C} & 58922 -- 59015 &  $\sim 500$ yr & $ \sim 2.5 \times 10^{14} \, \text{G}$  &  \(\sim  7 \times 10^{35} \ \mathrm{erg \, s^{-1}} \) \\
\cite{2021MNRAS.505.1311H} & 58936 -- 59092 & $\sim 500$ yr & $ \sim 2.5 \times 10^{14} \, \text{G}$  &  \(\sim  7 \times 10^{35} \ \mathrm{erg \, s^{-1}} \) \\
\cite{2022MNRAS.512.1687R} & 59117 – 59400 & $\sim 1000$ yr &  $ \sim 2 \times 10^{14} \, \text{G}$ & \(\sim  4 \times 10^{35} \ \mathrm{erg \, s^{-1}} \)  \\
This Work & 59520 – 59578  & $\sim 2500$ yr & $ \sim 1 \times 10^{14} \, \text{G}$  &  \(\sim  1 \times 10^{35} \ \mathrm{erg \, s^{-1}} \) \\
\bottomrule
\end{tabular}%
\tablecomments{The characteristic age ($\tau_c$) reported by \cite{2020ATel13559....1C} represents the first estimate for Swift~J1818.0$-$1607.  Subsequent values from \cite{2020MNRAS.498.6044C} and \cite{2021MNRAS.505.1311H} were inferred by fitting a linear function to the spin frequency measured at each epoch during their respective observing campaigns.  Estimates from \cite{2022MNRAS.512.1687R} and this work are derived from timing solutions. } 
\label{tab:charactersitic_age}%
\end{table*}%

\subsection{Flux Density \& Spectral Indices}

Our flux density measurements for Swift~J1818.0$-$1607 indicate a continuation of the declining trend observed around MJD 59300 (Figure~\ref{fig:flux_density_evolution}), with the source exhibiting a relatively stable, low-level radio emission state throughout our observing campaign. Placing Swift~J1818.0$-$1607 in the context of other radio magnetars can help clarify its potential evolutionary trajectory.

PSR~J1622$-$4950, the third discovered radio-loud magnetar, displayed a downward trend approximately two years after its discovery in 2009 \citep{2010ApJ...721L..33L}, before entering a low-flux regime by 2013, and stopped emitting any detectable radio emission entirely around five years post-discovery \citep{2017ApJ...841..126S}, before it was revived again in 2017 \citep{2018ApJ...856..180C}. A similar pattern was seen in XTE~J1810$-$197, which exhibited bright and highly variable radio emission \citep{2007ApJ...659L..37C, 2007ApJ...669..561C} after its 2003 outburst \citep{2004ApJ...609L..21I}, followed by a steady decline and eventual disappearance in late 2008 \citep{2016ApJ...820..110C}, before it got reactivated again in 2018 \citep{2019ApJ...874L..25G}. In contrast, 1E~1547.0$-$5408 was initially a bright and persistently detectable radio source after its 2007 discovery \citep{2007ApJ...667.1111G, 2007ApJ...666L..93C}. Following two X-ray outbursts in 2008 and 2009 \citep{2012ApJ...748....3D, 2012ApJ...748..133K}, its radio emission became markedly sporadic, with extended periods of non-detection lasting several months \citep{2009ATel.1907....1C, 2009ATel.1913....1B}. More recently, during another outburst in 2022, observations showed that the persistent radio emission from 1E~1547.0$-$5408 disappeared at least 22 days before the initial \textit{Swift}$-$BAT detection and was subsequently re-detected about two weeks later \citep{2023ApJ...945..153L}.

As of Febraury 2023, Swift~J1818.0$-$1607 continues to be detected in radio \citep{2024ApJ...971L..13D}, however continued long-term monitoring will be crucial to determine whether Swift~J1818.0$-$1607 eventually follows the fading and reactivation cycles seen in PSR~J1622$-$4950 and XTE~J1810$-$197, or transitions into a more sporadic and intermittent emission state like 1E1547.0$-$5408 or follows a unique evolutionary path. Our future work will shed light on the long-term evolution of Swift~J1818.0$-$1607  and its place within the broader population of radio magnetars.

Following its outburst, the pulsed radio spectrum of Swift~J1818.0$-$1607 was initially steep ($\alpha$  between $-\, 3.6$ and $-\, 1.8$) \citep{2020ATel13560....1M, 2020ApJ...896L..37L}, but gradually flattened over time \citep{2020ATel13649....1M}, with a transition from a steep spectrum  ($\alpha < -\, 1.48$, MJD 58936$-$58944) to a flat one  ($\alpha > -\, 0.63$, MJD 59015$–$59092) \citep{2021MNRAS.505.1311H, 2023MNRAS.523.2401B}.~\cite{2021MNRAS.502..127L} notes that this observed spectral flattening was driven by the emergence of a new profile component with an inverted spectrum, which persisted at high frequencies, while the original steep$-$spectrum component gradually faded and merged with it by MJD 59128.~Our observations show that Swift~J1818.0$-$1607 continues to exhibit a flat spectrum ($\alpha \gtrsim -1$), in contrast to the steep spectrum seen shortly after the outburst (Figure \ref{fig:spectral_index_evolution}).~  A similar spectral evolution was observed in PSR J1119$–$1627, a RPP that has exhibited magnetar-like outbursts \citep{2016ApJ...829L..21A, 2018MNRAS.480.3584D}. These characteristics align more closely with RPPs than with typical radio magnetars, leading to speculation that Swift~J1818.0$-$1607 may serve as an evolutionary link between the two populations \citep{2020ApJ...902....1H}.

Due to the limited bandwidth of these observations, we can not investigate the possibility of a spectral turnover at higher frequency, as seen by \cite{2021MNRAS.505.1311H}, where in one of their epochs (MJD 59076), they get an $\alpha$ = +0.5 below 32 GHz, whereas \cite{2020ATel14001....1T} reports $\alpha = -\, 1.4$ between 86 and 154 GHz at the same epoch.~Such turnover has also been observed in another magnetar XTE J1810$-$197. However, we note that the spectral turnover in XTE~J1810$-$197 occurs at substantially lower radio frequencies (MHz to a few GHz) compared to the tens of GHz turnover reported for Swift J1818.0$-$1607. As such, it remains unclear whether these turnovers are directly comparable, as they may originate from distinct physical processes.


\begin{figure}
    \centering
    \includegraphics[width=1.0\linewidth]{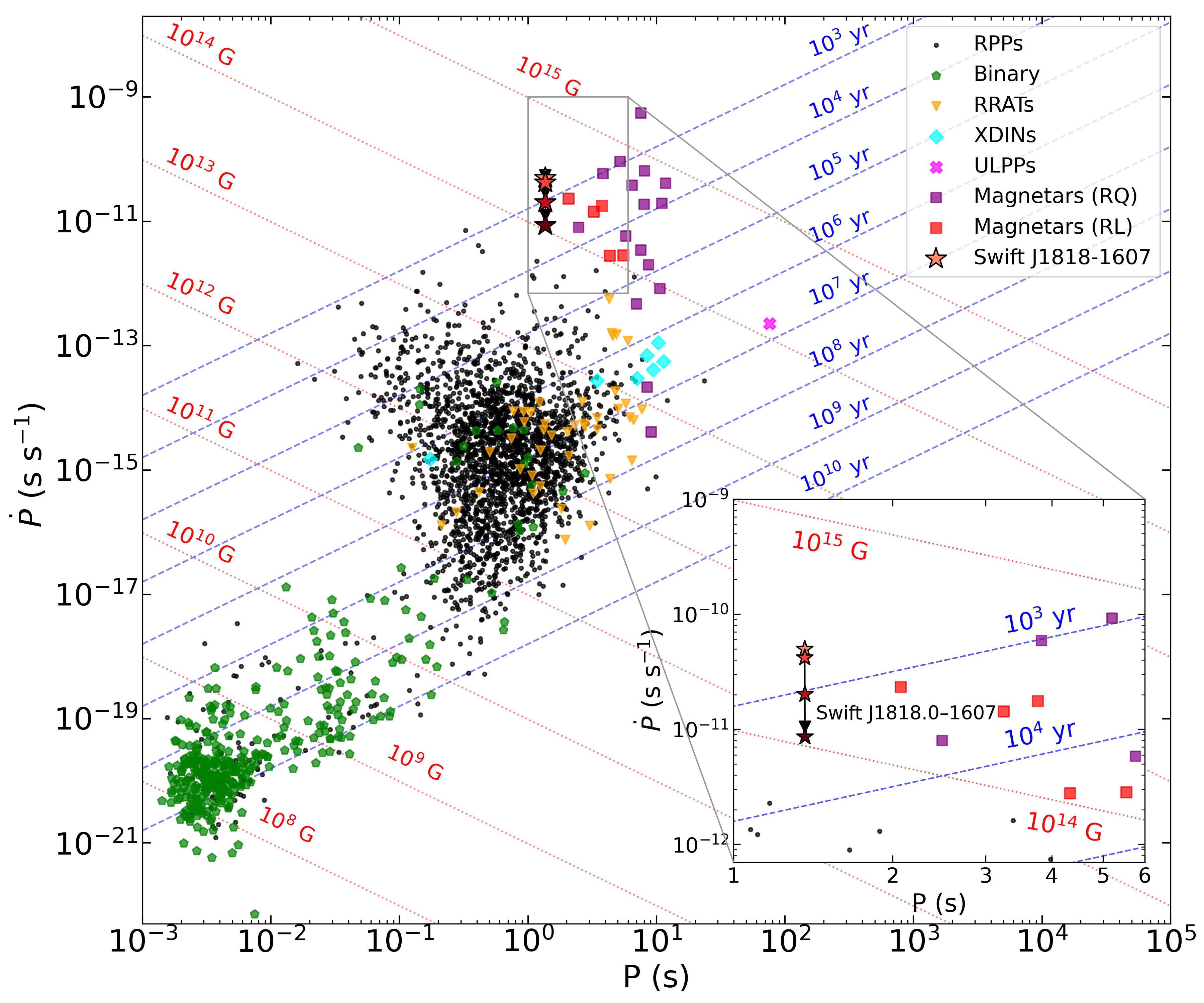}
    \caption{P$-\dot{P}$ diagram showing the position of Swift J1818.0$-$1607. The inset provides a zoomed$-$in view highlighting the temporal evolution of its spin period and spin$-$down rate across different epochs, with four data points (from top to bottom) corresponding to measurements reported by \cite{2020MNRAS.498.6044C}, \cite{2021MNRAS.505.1311H}, \cite{2022MNRAS.512.1687R}, and this work.}
    \label{fig:p_pdot_diagram}
\end{figure}

\vspace*{\fill}

\subsection{Inferred Properties of Swift~J1818.0$-$1607 on the P$-\dot{P}$ Diagram }

 Using the derived timing solution (Table \ref{tab:timing_solution}), we infer a  surface dipole magnetic field strength of $B \sim 1 \times 10^{14} \, \text{G}$, estimated using $B = 3.2 \times 10^{19} \sqrt{P\dot{P}}$ G, where $P$ is the spin period and $\dot{P}$ is its time derivative. This value is consistent with the typical range observed in the magnetar population \citep{2017ARA&A..55..261K}, but is  $\sim 3$ times lower than the value of $B \sim 3.5 \times 10^{14} \, \text{G}$, originally reported after the outburst \citep{2020ATel13559....1C}. Table \ref{tab:charactersitic_age} summarizes the evolution of key spin-down parameters for Swift~J1818.0$-$1607 across multiple observational campaigns \citep{2020ATel13559....1C, 2020MNRAS.498.6044C, 2021MNRAS.505.1311H, 2022MNRAS.512.1687R}, highlighting the decreasing trend in inferred magnetic field strength with time.

Figure \ref{fig:p_pdot_diagram} shows the position of Swift~J1818.0$-$1607 in the P -- $\dot{P}$,  highlighting its location within the broader pulsar and magnetar population. The inset provides a zoomed-in view of its temporal evolution, with four data points corresponding to P -- $\dot{P}$ measurements reported in previous studies \citep{2020MNRAS.498.6044C, 2021MNRAS.505.1311H, 2022MNRAS.512.1687R}, and this work. The downward trend in $\dot{P}$ over time reflects significant variability in the magnetar's spin-down rate, possibly driven by evolving magnetospheric conditions. This apparent movement in the P -- $\dot{P}$ diagram underscores the importance of continued long-term monitoring, to better understand the long-term spin-down properties of this magnetar.

Using the same timing solution, we infer a characteristic age of $\tau_c \sim 2500$ years for Swift~J1818.0$–$1607, calculated under the assumption of a constant braking index of $n = 3$, using the standard formula $\tau_c = P / 2\dot{P}$. This estimate is roughly an order of magnitude higher than the first reported age of $\sim 265$ years during the initial outburst phase \citep{2020ATel13559....1C}, and $\sim$ 2.5 times older than the most recent estimate ($\tau_{c} \sim 1000$ years, \citealt{2022MNRAS.512.1687R}). Table \ref{tab:charactersitic_age} summarizes the various age estimates for Swift~J1818.0$–$1607.~Taken together, these results suggest that the magnetar has effectively aged by $\sim$ 2000 years over the past two years, reflecting a significant and ongoing evolution in its spin$-$down behavior.

It is important to note that the characteristic age may not accurately reflect the true age of the magnetar. This estimate assumes a constant braking index of $n = 3$, which may not hold in practice, as the braking index can differ from this value or evolve over time \citep{2017MNRAS.467.3493J}. For instance, if the braking index is less than 3, as observed in a few magnetars \citep{2016MNRAS.456...55G}, it would result in a change in the magnetar's inferred age. Additionally, given that magnetars often exhibit irregular variations in their spin-down rate ($\dot{\nu}$), particularly during or shortly after outbursts, it can lead to a wide range of age estimates, as demonstrated in this study. Therefore, obtaining independent constraints on the age of Swift~J1818.0$–$1607 is crucial. While no coincident supernova remnant (SNR) has previously been identified to directly constrain the magnetar's age \citep{2025JApA...46...14G, 2012AdSpR..49.1313F}, recent radio observations have revealed diffuse radio emission around Swift~J1818.0$–$1607 \citep{2023ApJ...943...20I}, suggesting that it originates from the shell of the supernova remnant where the magnetar was formed. This hypothesis is further corroborated by recent VLBI observations, which measured the magnetar's 3$-$D velocity to be 190 km s$^{-1}$ \citep{2024ApJ...971L..13D}, indicating that the source has likely not traveled far from where it was born. High-resolution follow-up radio observations are essential to robustly characterize the diffuse emission and determine whether its spectral and morphological properties are consistent with an SNR shell.


\section{Summary} 
\label{sec:summaryandconclusions}


We carried out observations of the radio magnetar Swift~J1818.0$–$1607 with the GBT at a central frequency of 2.0 GHz in November 2021, approximately 20 months after its March 2020 outburst.~From these observations, we derived a phase-connected timing solution spanning the duration of our monitoring campaign.

Using our phase-connected timing solution, we infer a characteristic age of $\sim 2500$ years for Swift~J1818.0$–$1607, significantly older than previous estimates (see Table \ref{tab:charactersitic_age}). This increase reflects the continued evolution of the magnetar’s spin-down rate, and underscores the limitations of inferring its $\tau_{c}$ from P $-$ $\dot{P}$ measurements, particularly for young, active magnetars with highly variable spin-down behavior.

Throghout our observing campaign, Swift~J1818.0$–$1607 exhibited a stable pulse profile, characterized by a single, narrow peak ($W_{50} \sim 18-23$ ms) at 2.0 GHz with a small precursor component and no evidence of a postcursor, and maintaned a magnetar-like flat radio spectrum. Our analysis suggests that the magnetar likely underwent a mode$-$switching episode during the $\sim$100-day gap (MJD 59426 $–$ 59520) preceding our observations, as indicated by changes in the pulse profile, spin$-$down rate, and flux density relative to the emission mode (Mode 4) reported in earlier epochs \citep{2022MNRAS.512.1687R}.~The observed variability in the spin-down rate of Swift~J1818.0$–$1607 is not surpising, as similar behavior has been observed in other magnetars and high B-field pulsars \citep{2012ApJ...748....3D, 2012ApJ...748..133K, 2015ApJ...800...33A, 2016ApJ...829L..21A, 2016ApJ...820..110C, 2017ApJ...841..126S, 2019ApJ...874L..14D, 2019MNRAS.488.5251L, 2023ApJ...945..153L, 2023ApJ...956...93H}.

In future work, we plan on presenting a comprehensive analysis of multi-frequency data from the rest of our monitoring campaign covering the period from MJD 59700 to MJD 60300. Our primary focus will be on the long-term evolution of the pulsed radio emission from Swift~J1818.0$–$1607 in both frequency and time, by leveraging contemporaneous multi-frequency observations in the L and C bands across most epochs.

\section*{Orcid IDs}
\begin{tabbing}
\hspace{3cm} \= \hspace{2cm} \= \kill
\text{Moaz Abdelmaguid:} \> \orcidlink{0000-0002-4441-7081}\href{https://orcid.org/0000-0002-4441-7081}{0000-0002-4441-7081} \\
\text{Paulo Freire:} \> \orcidlink{0000-0003-1307-9435}\href{https://orcid.org/0000-0003-1307-9435}{0000-0003-1307-9435} \\
\text{Joseph  Gelfand:} \> \orcidlink{0000-0003-4679-1058}\href{https://orcid.org/0000-0003-4679-1058}{0000-0003-4679-1058} \\
\text{Yogesh Maan:} \> \orcidlink{0000-0002-0862-6062}\href{https://orcid.org/0000-0002-0862-6062}{0000-0002-0862-6062} \\
\text{Samayra Straal:} \> \orcidlink{0000-0003-4136-7848}\href{https://orcid.org/0000-0003-4136-7848}{0000-0003-4136-7848} \\
\text{J. A. J. Alford:} \> \orcidlink{0000-0002-2312-8539}\href{https://orcid.org/0000-0002-2312-8539}{0000-0002-2312-8539}
\end{tabbing}

\vspace*{\fill}

\section*{Acknowledgements}

The authors would like to thank the anonymous referee for the comments and suggestions that helped make this manuscript better.

M.A. is supported by a graduate research assistanship at NYU Abu Dhabi, which is funded by the Executive Affairs Authority of the Emirate of Abu Dhabi, as adminstrated by Tamkeen. The research of J.D.G is funded by NYUAD grant AD 022. Both M.A. and J.D.G. received support form the NYU Abu Dhabi Research Institute grant to the CASS. The National Radio Astronomy Observatory and Green Bank Observatory are facilities of the U.S. National Science Foundation operated under cooperative agreement by Associated Universities.

M.A. would like to thank Scott M. Ransom and Ryan S. Lynch for answering questions related to PRESTO. M.A. would also like to thank Zhi-Peng Huang and Kaustubh Rajwade for sending their data which has been used in this publication. 

\newpage

\section*{Facilities \& Softwares}

\emph{Facilities}: GBT \\
\emph{Software}: \texttt{TEMPO2} \citep{2006ChJAS...6b.189H}, \\ \texttt{PRESTO} \citep{2002AJ....124.1788R}, \texttt{DSPSR} \citep{2011PASA...28....1V},  \texttt{PSRCHIVE} \citep{2004PASA...21..302H} \& \texttt{Scipy} \citep{2020NatMe..17..261V}.

\section*{Data Availability}

The data underlying Figure~\ref{fig:residuals} is listed in Table~\ref{tab:toas_measurements} in the Appendix. Those underlying Figures~\ref{fig:flux_density_evolution} and~\ref{fig:spectral_index_evolution} are provided in Table~\ref{tab:fluxdensity_pulse_width}. Additionally, the data for Figure~\ref{fig:spin_frequency_slopes} is presented in Table~\ref{tab:spin_frequency_evolution} in the Appendix. Finally, the data tracing the trajectory of Swift~J1818.0$-$1607 on the $P$–$\dot{P}$ diagram in Figure~\ref{fig:p_pdot_diagram} is listed in Table~\ref{tab:charactersitic_age}.

\appendix
\renewcommand{\thetable}{A.\arabic{table}}
\setcounter{table}{0}


In Table~\ref{tab:toas_measurements}, we provide the full set of extracted time-of-arrival (ToA) measurements used to construct the phase-connected timing solution in Table~\ref{tab:timing_solution}. Their timing residuals are shown in Figure~\ref{fig:residuals}. Each row lists the observatory code, frequency, MJD of the ToA, and the associated uncertainty in microseconds. Table~\ref{tab:spin_frequency_evolution} lists the spin frequency measurements obtained from our observing campaign, which were used in Figure~\ref{fig:spin_frequency_slopes}. Each measurement corresponds to a single epoch, with $\sigma_\nu$ denoting the statistical uncertainty associated with each spin frequency measurement.

\small
\begin{longtable}{cccc}
\caption{Time of arrival (ToA) measurements for Swift~J1818.0$-$1607 taken with the GBT. Each row lists the observatory code (GBT = 1), observing frequency (MHz), MJD of the ToA, and the uncertainty in $\mu$s. These data were used to construct the phase-connected timing solution listed in Table~\ref{tab:timing_solution}, and whose residuals are shown in Figure~\ref{fig:residuals}.}
\label{tab:toas_measurements} \\
\toprule
Obs. Code & Frequency (MHz) & MJD & Uncertainty ($\mu$s) \\
\midrule
\endfirsthead

\toprule
Obs. Code & Frequency (MHz) & MJD & Uncertainty ($\mu$s) \\
\midrule
\endhead

\bottomrule
\endfoot

1 & 2000.195 & 59520.7650042251161 & 360.52 \\
1 & 2000.195 & 59520.7660469857919 & 320.79 \\
1 & 2000.195 & 59520.7670739406895 & 382.48 \\
1 & 2000.195 & 59520.7681008908125 & 319.86 \\
1 & 2000.195 & 59520.7711975653497 & 334.83 \\
1 & 2000.195 & 59536.7135290693987 & 353.97 \\
1 & 2000.195 & 59536.7152511941632 & 356.62 \\
1 & 2000.195 & 59536.7169733116307 & 338.93 \\
1 & 2000.195 & 59536.7204017445736 & 339.84 \\
1 & 2000.195 & 59550.8868235680141 & 311.14 \\
1 & 2000.195 & 59550.8885456768594 & 309.47 \\
1 & 2000.195 & 59550.8868235630431 & 329.95 \\
1 & 2000.195 & 59550.8888932390432 & 443.79 \\
1 & 2000.195 & 59559.8345127531953 & 407.93 \\
1 & 2000.195 & 59559.8365824132791 & 366.77 \\
1 & 2000.195 & 59559.8376093264854 & 417.73 \\
1 & 2000.195 & 59559.8386362598791 & 542.49 \\
1 & 2000.195 & 59559.8407059098786 & 321.52 \\
1 & 2000.195 & 59559.8427755654863 & 410.58 \\
1 & 2000.195 & 59559.8438024861103 & 425.01 \\
1 & 2000.195 & 59564.8013005036540 & 401.33 \\
1 & 2000.195 & 59564.8043970641237 & 402.85 \\
1 & 2000.195 & 59564.8085205477038 & 370.73 \\
1 & 2000.195 & 59564.8095474593230 & 380.80 \\
1 & 2000.195 & 59568.7141911149129 & 420.68 \\
1 & 2000.195 & 59568.7148862444787 & 440.61 \\
1 & 2000.195 & 59568.7169558566322 & 408.80 \\
1 & 2000.195 & 59568.7453303473052 & 343.72 \\
1 & 2000.195 & 59572.6262417115981 & 768.09 \\
1 & 2000.195 & 59572.6272686235020 & 465.28 \\
1 & 2000.195 & 59572.6293382250276 & 370.71 \\
1 & 2000.195 & 59572.6313920679945 & 683.49 \\
1 & 2000.195 & 59576.5999669558361 & 453.98 \\
1 & 2000.195 & 59576.6009938571775 & 363.49 \\
1 & 2000.195 & 59576.6040903807802 & 337.17 \\
1 & 2000.195 & 59577.7544138512075 & 341.59 \\
1 & 2000.195 & 59577.7561358842822 & 317.18 \\
1 & 2000.195 & 59577.7578579233058 & 437.16 \\
1 & 2000.195 & 59577.7595799644018 & 396.98 \\
1 & 2000.195 & 59577.7818085152489 & 537.15 \\
1 & 2000.195 & 59577.7859477464216 & 325.39 \\
1 & 2000.195 & 59577.7869746523945 & 396.94 \\
1 & 2000.195 & 59577.7910980569357 & 444.58 \\
1 & 2000.195 & 59578.8012713277590 & 303.80 \\
1 & 2000.195 & 59578.8029933650074 & 358.54 \\
1 & 2000.195 & 59578.8064374293458 & 376.16 \\
1 & 2000.195 & 59578.8081436699751 & 569.32 \\
\end{longtable}

\begin{table*}[htbp]
\centering
\caption{Pulse width, phase, and amplitude measurements for Swift J1818.0$-$1607 during our monitoring campaign at 2.0 GHz.}
\begin{tabular}{ccccccc}
\toprule
\toprule
MJD & \multicolumn{3}{c}{Main Component} & \multicolumn{3}{c}{Pre-cursor Component} \\
- & Width (ms) & Phase & Amplitude & Width (ms) & Phase & Amplitude \\
\midrule
59520.76 & 17.3(5) & 0.375 & 0.81 & 48.0(3) & 0.363& 0.20 \\
59536.71 & 18.6(8) & 0.374 & 0.94 & 41.3(6) & 0.359 & 0.12\\
59550.88 & 17.3(5) & 0.374  & 0.84 & 41.3(6) & 0.365 & 0.18 \\
59559.84 & 17.3(5) & 0.375 & 0.86 & 34.6(9) & 0.366 & 0.21 \\
59564.80 & 20.0(1) & 0.375 & 0.90 & 36.0(3) & 0.362 & 0.21 \\
59568.72 & 20.0(3) & 0.374 & 0.853 & 36.0(4) & 0.364 & 0.17 \\ 
59572.63 & 14.6(8) & 0.376 & 0.72 & 37.3(6) & 0.367 & 0.33 \\
59576.59 & 17.3(5) & 0.374 & 0.85 & 41.3(6) & 0.364 & 0.18 \\
59577.76 & 20.0(1) & 0.374 & 0.83 & 41.3(6) & 0.364 & 0.20 \\
59578.80 & 18.6(8) & 0.374 & 0.87 & 44.0(3) & 0.366 & 0.16 \\
\bottomrule
\end{tabular}%
\tablecomments{Gaussian fit parameters of the main and pre-cursor pulse components of Swift J1818.0$-$1607 observed at 2.0\,GHz during our monitoring campaign. For each epoch, we report the full-width at half-maximum (FWHM) in ms, the fitted pulse phase, and the normalized amplitude of each component. Uncertainties are quoted at the $1\sigma$ level in parentheses and reflect the uncertainty in the last digit(s).}
\label{tab:gauss_fit_parameters}%
\end{table*}

\begin{table}[ht]
\begin{center}
\caption{Spin frequency measurements of Swift~J1818.0$-$1607 from our observing campaign. $\sigma_\nu$ represents the statistical uncertainty associated with each spin frequency measurement. These data were used to create Figure~\ref{fig:spin_frequency_slopes}.}
\label{tab:spin_frequency_evolution}
\begin{tabular}{ccc}
\hline
\hline
MJD & $\nu$ (Hz) & $\sigma_\nu$ (Hz) \\
\hline
59520.76339 & 0.7326252354 & $6.07 \times 10^{-7}$ \\
59536.71396 & 0.7326148968 & $5.77 \times 10^{-7}$ \\
59550.88215 & 0.7326072100 & $3.94 \times 10^{-7}$ \\
59559.83677 & 0.7326073648 & $4.17 \times 10^{-7}$ \\
59564.79993 & 0.7326066743 & $4.96 \times 10^{-7}$ \\
59568.70900 & 0.7326099055 & $7.07 \times 10^{-7}$ \\
59568.73670 & 0.7325946103 & $1.16 \times 10^{-6}$ \\
59572.62728 & 0.7325980647 & $4.81 \times 10^{-7}$ \\
59576.59604 & 0.7326033346 & $8.00 \times 10^{-7}$ \\
59577.74870 & 0.7326008607 & $8.17 \times 10^{-7}$ \\
59577.77600 & 0.7325999324 & $6.20 \times 10^{-7}$ \\
59578.79600 & 0.7326012217 & $6.29 \times 10^{-7}$ \\
\hline
\end{tabular}
\end{center}
\end{table}

\newpage

\bibliography{Manuscript}{}
\bibliographystyle{aasjournal}

\end{document}